\documentclass{iopart}

\usepackage{xspace}

\usepackage{iopams}

\usepackage{hyperref}
\usepackage[all]{hypcap}

\def\DDbar   {\ensuremath{\kern -0.1em \stackrel{\kern 0.1em \textsf{\fontsize{5pt}{1em}\selectfont(---)}}{D}\kern -0.3em}\xspace}
\def\AAbar   {\ensuremath{\kern -0.2em \stackrel{\kern 0.2em \textsf{\fontsize{5pt}{1em}\selectfont(---)}}{A}\kern -0.3em}\xspace}
\def\AfAfbar   {\ensuremath{\kern -0.2em \stackrel{\kern 0.2em \textsf{\fontsize{5pt}{1em}\selectfont(---)}}{A}_{\kern -0.3em f}\kern -0.3em}\xspace}
\def\dacp     {\ensuremath{\Delta A_{\CP}}\xspace}
\def\CP                {\ensuremath{C\!P}\xspace}
\def\Dbar    {\kern 0.2em\overline{\kern -0.2em \PD}{}\xspace}
\def\D       {\ensuremath{\PD}\xspace}

\def\Dz      {\ensuremath{\D^0}\xspace}
\def\Dzb     {\ensuremath{\Dbar^0}\xspace}
\def\DzDzb   {\ensuremath{\Dz {\kern -0.16em \Dzb}}\xspace}
\def\Dp      {\ensuremath{\D^+}\xspace}
\def\Dm      {\ensuremath{\D^-}\xspace}

\def\DpDm    {\ensuremath{\Dp {\kern -0.16em \Dm}}\xspace}

\def\lhcb {LHCb\xspace}

\def\cdf    {CDF\xspace}
\def\PD      {\ensuremath{D}\xspace}                 
\def\PK      {\ensuremath{K}\xspace}                 
\def\kaon  {\ensuremath{\PK}\xspace}
\def\Kp    {\ensuremath{\kaon^+}\xspace}
\def\Km    {\ensuremath{\kaon^-}\xspace}
\def\Ppi         {\ensuremath{\pi}\xspace}                 
\def\pion  {\ensuremath{\Ppi}\xspace}
\def\pip   {\ensuremath{\pion^+}\xspace}
\def\pim   {\ensuremath{\pion^-}\xspace}
\def\ycp        {\ensuremath{y_{CP}}\xspace}
\def\agamma     {\ensuremath{A_{\Gamma}}\xspace}
\def\KS    {\ensuremath{\kaon^0_{\rm\scriptstyle S}}\xspace} 
\newcommand{\ie}{\mbox{\itshape i.e.}}
\def\PB      {\ensuremath{\mathrm{B}}\xspace}                 
\def\Bbar    {\kern 0.2em\overline{\kern -0.2em \PB}{}\xspace}

\begin{document}

\title{On the interplay of direct and indirect \CP violation in the charm sector}

\author{M Gersabeck$^1$, M Alexander$^2$, S Borghi$^{2,3}$, VV Gligorov$^1$ and C Parkes$^{3,1}$}
\address{$^1$ European Organization for Nuclear Research (CERN), Geneva, Switzerland}
\address{$^2$ School of Physics and Astronomy, University of Glasgow, Glasgow, United Kingdom}
\address{$^3$ School of Physics and Astronomy, University of Manchester, Manchester, United Kingdom}
\ead{marco.gersabeck@cern.ch}

\begin{abstract}
  \noindent
  Charm mixing and \CP violation observables are examined in the light of the recently reported evidence from \lhcb for \CP violation in the charm sector.
  If the result is confirmed as being due to direct \CP violation at the $1\%$ level, its effect will need to be taken into account in the interpretation of \CP violation observables.
  The contributions of direct and indirect \CP violation to the decay rate asymmetry difference \dacp and the ratios of effective lifetimes \agamma and \ycp are considered here. 
  Terms relevant to the interpretation of future high precision measurements which have been neglected in previous literature are identified.
\end{abstract}

\pacs{13.25.Ft} 
\submitto{\jpg}

\maketitle

\noindent
Charm, after strange and beauty, is the last system of neutral flavoured mesons where \CP violation remains to be discovered.
While neutral \PB mesons are characterised by their mass splitting leading to fast oscillations and neutral kaons by their width splitting which results in a short-lived and a long-lived state, neutral charm mesons have a very small splitting in both mass and width.
For charm, compared to the beauty sector, this leads to rather subtle mixing-related effects in time-dependent as well as in time-integrated charm measurements, which are examined in detail here.

First evidence for \CP violation in the charm sector has recently been reported by the \lhcb collaboration in the study of the difference of the time-integrated asymmetries of $\Dz\to\Kp\Km$ and $\Dz\to\pip\pim$ decay rates through the parameter \dacp~\cite{Aaij:2011in}.
This measurement is primarily sensitive to the difference in direct \CP violation between the two final states as discussed further below.
Direct \CP violation depends on the final state and is the asymmetry of the rates of particle and antiparticle decays.
It can be caused by a difference in the magnitude of the decay rates or by a difference in their phase.
Indirect \CP violation is considered universal, \ie\ final-state independent, and is an asymmetry in the mixing rate or in its weak phase.

Indirect \CP violation can be measured in time-dependent analyses.
To date, two types of measurements were used to search for indirect \CP violation in the charm sector.
One uses the asymmetry of the lifetimes, \agamma, measured in \Dz and \Dzb decays to the \CP eigenstates $\Kp\Km$ or $\pip\pim$~\cite{Staric:2007dt,Aubert:2009ck,Aaij:2011ad}.
The other is a time-dependent Dalitz plot analysis of \Dz and \Dzb decays to $\KS\pip\pim$ or $\KS\Kp\Km$~\cite{Abe:2007rd,delAmoSanchez:2010xz}.
Another observable studied related to \agamma is \ycp, which is given by the deviation from one of the ratio of the lifetimes measured in decays to a Cabibbo-allowed, \CP averaged, and a Cabibbo-suppressed, \CP eigenstate, final state.
Any deviation of a measurement of \ycp from that of the mixing parameter $y$ would signal \CP violation.

In the interpretation of \agamma and \ycp, direct \CP violation is commonly neglected~\cite{Bergmann:2000id}.
In the light of the new evidence this assumption is no longer justified.
The relevance of direct \CP violation to a measurement of \agamma has previously been pointed out in~\cite{Kagan:2009gb}.
However, a closer look at both \agamma and \ycp is necessary to examine the contribution of direct and indirect \CP violation in these observables as well as their connection to \dacp.

The mass eigenstates of neutral \PD mesons, $|D_{1,2}\rangle$, with masses $m_{1,2}$ and widths $\Gamma_{1,2}$ can be written as linear combinations of the flavour eigenstates $|D_{1,2}\rangle=p|\Dz\rangle\pm{}q|\Dzb\rangle$, with complex coefficients $p$ and $q$ which satisfy $|p|^2+|q|^2=1$.
The average mass and width are defined as $m\equiv(m_1+m_2)/2$ and $\Gamma\equiv(\Gamma_1+\Gamma_2)/2$.
The \PD mixing parameters are defined using the mass and width difference as $x\equiv(m_2-m_1)/\Gamma$ and $y\equiv(\Gamma_2-\Gamma_1)/2\Gamma$.
The phase convention of $p$ and $q$ is chosen such that $\CP|\Dz\rangle=-|\Dzb\rangle$.

According to~\cite{Kagan:2009gb}, the time dependent decay rates of \Dz and \Dzb decays to the final state $f$, which is a \CP eigenstate with eigenvalue $\eta_{\CP}$, can be expressed as
\begin{eqnarray}
\label{eqn:start}
\fl\Gamma(\Dz(t)\to f)=\frac{1}{2}\rme^{-\tau}\left|A_f\right|^2 \Big\{ & \left(1+|\lambda_f|^2\right)\cosh(y\tau)+\left(1-|\lambda_f|^2\right)\cos(x\tau) \nonumber\\
 &+2\Re(\lambda_f)\sinh(y\tau)-2\Im(\lambda_f)\sin(x\tau)\Big\},\nonumber\\
\fl\Gamma(\Dzb(t)\to f)=\frac{1}{2}\rme^{-\tau}\left|\bar{A}_f\right|^2 \Big\{ & \left(1+|\lambda^{-1}_f|^2\right)\cosh(y\tau)+\left(1-|\lambda^{-1}_f|^2\right)\cos(x\tau) \nonumber\\
 &+2\Re(\lambda^{-1}_f)\sinh(y\tau)-2\Im(\lambda^{-1}_f)\sin(x\tau)\Big\},
\end{eqnarray}
where $\tau\equiv\Gamma t$, \AfAfbar~~are the decay amplitudes and $\lambda_f$ is given by
\begin{equation}
\lambda_f=\frac{q\bar{A}_f}{pA_f}=-\eta_{\CP}\left|\frac{q}{p}\right|\left|\frac{\bar{A}_f}{A_f}\right|\rme^{i\phi},
\end{equation}
where $\eta_{\CP}$ is the \CP eigenvalue of the final state $f$ and $\phi$ is the \CP violating relative phase between $q/p$ and $\bar{A}_f/A_f$.
Introducing $|q/p|^{\pm 2}\approx1\pm A_m$ and $|\bar{A}_f/A_f|^{\pm 2}\approx1\pm A_d$, one can write
\begin{equation}
\label{eqn:lambda_approx}
|\lambda_f^{\pm1}|^2\approx(1\pm A_m)(1\pm A_d),
\end{equation}
where $A_m$ represents a \CP violation contribution from mixing and $A_d$ from direct \CP violation and where both $A_m$ and $A_d$ are assumed to be small.

Expanding~\eref{eqn:start} up to second order in $\tau$ one can write the effective lifetimes, \ie\ those measured as a single exponential, as
\begin{eqnarray}
\fl\hat{\Gamma} (\DDbar(t)\to f)\approx & \Gamma\Bigg\{ 1 + \left[ 1 \pm\frac{1}{2}(A_m+A_d)-\frac{1}{8}(A_m^2-2A_mA_d)\right]\eta_{\CP}(y\cos\phi\mp x\sin\phi)\nonumber\\
& \quad\mp A_m(x^2+y^2)\pm 2A_my^2\cos^2\phi\mp 4xy\cos\phi\sin\phi\Bigg\},
\end{eqnarray}
where terms below order $10^{-5}$ have been ignored.
The experimental constraints~\cite{Asner:2010qj} give $x$, $y$, and $A_d$ for the final states $\Kp\Km$ and $\pip\pim$ of order $10^{-2}$, and $A_m$ and $\sin\phi$ of order $10^{-1}$.
The sum of measurements of \Dz and \Dzb decays leads to the definition of the observable \ycp which is given by
\begin{equation}
\fl\ycp=\frac{\hat{\Gamma} + \hat{\bar{\Gamma}}}{2\Gamma} -1 \approx \eta_{\CP}\Bigg\{\left[ 1 -\frac{1}{8}(A_m^2-2A_mA_d)\right]y\cos\phi -\frac{1}{2}(A_m+A_d)x\sin\phi\Bigg\}.
\end{equation}
The difference of measurements of \Dz and \Dzb decays leads to the parameter \agamma which is defined as
\begin{eqnarray}
\fl\agamma=(\hat{\Gamma} - \hat{\bar{\Gamma}})(\hat{\Gamma} + \hat{\bar{\Gamma}})^{-1} \approx & \bigg[\frac{1}{2}(A_m+A_d)y\cos\phi-\left(1-\frac{1}{8}A_m^2\right)x\sin\phi- A_m(x^2+y^2) \nonumber\\
& + 2A_my^2\cos^2\phi - 4xy\cos\phi\sin\phi \bigg]\frac{\eta_{\CP}}{1+\ycp}.
\end{eqnarray}
The weak phase $\phi$ has not been assumed to be universal.
When averaging measurements from different channels, a potential decay-dependent weak phase of the amplitude ratio has to be taken into account~\cite{Kagan:2009gb}.
Expanding only up to order $10^{-4}$ leads to
\begin{equation}
\label{eqn:ycp}
\ycp \approx \eta_{\CP}\left[\left( 1 -\frac{1}{8}A_m^2\right)y\cos\phi -\frac{1}{2}(A_m)x\sin\phi\right],
\end{equation}
and
\begin{equation}
\label{eqn:agamma}
\fl\agamma \approx \bigg[\frac{1}{2}(A_m+A_d)y\cos\phi-x\sin\phi \bigg]\frac{\eta_{\CP}}{1+\ycp} \approx \eta_{\CP}\left[\frac{1}{2}(A_m+A_d)y\cos\phi-x\sin\phi\right].
\end{equation}
The difference of \ycp evaluated in \eref{eqn:ycp} from the expression used in literature~\cite{Bergmann:2000id} so far is the term $\eta_{\CP}\frac{1}{8}A_m^2y\cos\phi$, which can be of similar order as $\frac{1}{2}A_mx\sin\phi$ and should therefore not be ignored.
\Eref{eqn:agamma} shows that there can be a significant contribution to \agamma from direct \CP violation.
Assuming $y=1\%$ and $\cos\phi=1$, direct \CP violation at the level of $A_d/2=1\%$ would lead to a contribution to \agamma of $10^{-4}$.
Current measurements yield a sensitivity of a few $10^{-3}$~\cite{Staric:2007dt,Aubert:2009ck,Aaij:2011ad}.
Future measurements at \lhcb and future \PB factory experiments are expected to reach uncertainties of the level of $10^{-4}$, \ie\ that of the direct \CP violation contribution.
More precise measurements may well change the approximations made in \eref{eqn:ycp} and \eref{eqn:agamma}, in particular a measurement of $A_m\lesssim A_d$.

In time-integrated measurements the rate asymmetry is measured which is defined as
\begin{equation}
\label{eqn:acp}
A_{\CP}\equiv\frac{\Gamma(\Dz\to f)-\Gamma(\Dzb\to f)}{\Gamma(\Dz\to f)+\Gamma(\Dzb\to f)}.
\end{equation}
Introducing
\begin{equation}
a_{\CP}^{dir}\equiv\frac{|A_f|^2-|\bar{A}_f|^2}{|A_f|^2+|\bar{A}_f|^2}=\frac{1-\left|\frac{\bar{A}_f}{A_f}\right|^2}{1+\left|\frac{\bar{A}_f}{A_f}\right|^2}=\frac{-A_d}{2+A_d}\approx-\frac{1}{2}A_d,
\end{equation}
and using \eref{eqn:start}, \eref{eqn:acp} becomes
\begin{equation}
A_{\CP}\approx a_{\CP}^{dir}-\agamma(1-(a_{\CP}^{dir})^2)\frac{\langle t \rangle}{\tau}\approx a_{\CP}^{dir}-\agamma\frac{\langle t \rangle}{\tau},
\end{equation}
where $\langle t \rangle$ denotes the average decay time of the observed candidates.
Terms in $\langle t \rangle^2$ are below order $10^{-4}$, given current experimental constraints, and have been ignored.

A common way to reduce experimental systematic uncertainties is to measure the difference in time-integrated asymmetries in related final states.
For the two-body final states $\Kp\Km$ and $\pip\pim$, this difference is given by
\begin{eqnarray}
\dacp & \equiv A_{\CP}(\Kp\Km)-A_{\CP}(\pip\pim)\nonumber\\
& = a_{\CP}^{dir}(\Kp\Km)-a_{\CP}^{dir}(\pip\pim)\nonumber\\
& \quad- \agamma(\Kp\Km)\frac{\langle t(\Kp\Km) \rangle}{\tau} + \agamma(\pip\pim)\frac{\langle t(\pip\pim) \rangle}{\tau}.
\end{eqnarray}
Assuming the \CP violating phase $\phi$ to be universal~\cite{Grossman:2006jg} this can be rewritten as
\begin{eqnarray}
\dacp \approx \Delta a_{\CP}^{dir}\left(1+y\cos\phi\frac{\overline{\langle t \rangle}}{\tau}\right) + \left(a_{\CP}^{ind}+\overline{a_{\CP}^{dir}}y\cos\phi\right)\frac{\Delta\langle t \rangle}{\tau}
\end{eqnarray}
where $\Delta X \equiv X(\Kp\Km)-X(\pip\pim)$, $\Delta X \equiv X(\Kp\Km)-X(\pip\pim)$, and $a_{\CP}^{ind}=-(A_m/2)y\cos\phi+x\sin\phi$.
The ratio $\overline{\langle t \rangle}/\tau$ is equal to one for the lifetime-unbiased \PB factory measurements~\cite{Aubert:2007if, Staric:2008rx} and is $2.083\pm0.001$ for \lhcb~\cite{Aaij:2011in} and $2.53\pm0.02$ for \cdf~\cite{Aaltonen:2011zzz}, thus leading to a correction of $\Delta a_{\CP}^{dir}$ of the order of $10^{-2}$.
The factor $\Delta\langle t \rangle/\tau$ multiplying the indirect \CP violation is zero for the \PB factory measurements and ranges from $0.098\pm0.003$ to $0.26\pm0.01$ for \lhcb and \cdf, respectively.
Therefore, \dacp is largely a measure of direct \CP violation while an obvious contribution from indirect \CP violation exists.
The contribution from direct \CP violation to \agamma pointed out in \eref{eqn:agamma} leads to a term proportional to $y$.
This term may be of similar size as the term proportional to $\Delta\langle t \rangle$ and should therefore be taken into account.

In summary, the mixing and \CP violation parameters \ycp, \agamma and \dacp have been discussed in the light of the recent evidence for \CP violation in the \Dz sector.
The parameter \ycp is least affected by direct \CP violation, however, it contains a term which has been neglected in the literature so far and which can be of the same order as the constribution proportional to $x$.
A measurement of \agamma can exhibit a contribution of direct \CP violation at the level of $10^{-4}$, comparable to the expected future experimental sensitivity.
The direct \CP violation term in the \dacp measurement contains a contribution proportional to $y$.
The interpretation of future high precision measurements of these observables will need to take account of these contributions.

\ack
The authors would like to thank Tim Gershon, Bostjan Golob, Alex Kagan and Vincenzo Vagnoni for helpful discussions and comments.
Furthermore, the authors thank the \lhcb collaboration whose work inspired this paper.
MG and VVG are supported by a Marie Curie Action: ``Cofunding of the CERN Fellowship Programme (COFUND-CERN)'' of the European Community's Seventh Framework Programme under contract number (PCOFUND-GA-2008-229600). MA, SB and CP acknowledge the support of the STFC (United Kingdom).

\section*{References}
\bibliographystyle{unsrt}
\bibliography{main}

\begin{thebibliography}{10}

\bibitem{Aaij:2011in}
Aaij~R {\it et al.}~[\lhcb collaboration].
\newblock {Evidence for \CP violation in time-integrated $\Dz\to h^-h^+$ decay
  rates}.
\newblock 2011.
\newblock {\it Preprint hep-ex 1112.0938}.

\bibitem{Staric:2007dt}
Staric~M {\it et al.}~[\belle collaboration].
\newblock {Evidence for \Dz - \Dzb Mixing}.
\newblock {\em Phys. Rev. Lett.}, 98:211803, 2007.

\bibitem{Aubert:2009ck}
Aubert~B {\it et al.}~[\babar collaboration].
\newblock {Measurement of \Dz - \Dzb Mixing using the Ratio of Lifetimes for
  the Decays $\Dz\to\Km\pip$ and $\Kp\Km$}.
\newblock {\em Phys. Rev.}, D80:071103, 2009.

\bibitem{Aaij:2011ad}
Aaij~R {\it et al.}~[\lhcb collaboration].
\newblock {Measurement of mixing and \CP violation parameters in two-body charm
  decays}.
\newblock 2011.
\newblock {\it Preprint hep-ex 1112.4698}.

\bibitem{Abe:2007rd}
Abe~K {\it et al.}~[\belle collaboration].
\newblock {Measurement of \Dz-\Dzb Mixing Parameters in $\Dz\to\KS\pip\pim$
  decays}.
\newblock {\em Phys. Rev. Lett.}, 99:131803, 2007.

\bibitem{delAmoSanchez:2010xz}
Del Amo Sanchez~P {\it et al.}~[\babar collaboration].
\newblock {Measurement of $\Dz$-$\Dzb$ mixing parameters using
  $\Dz\to\KS\pip\pim$ and $\Dz\to\KS\Kp\Km$ decays}.
\newblock {\em Phys. Rev. Lett.}, 105:081803, 2010.

\bibitem{Bergmann:2000id}
Bergmann S, Grossman Y, Ligeti Z, Nir Y, and Petrov~A A.
\newblock {Lessons from CLEO and FOCUS measurements of \Dz-\Dzb mixing
  parameters}.
\newblock {\em Phys. Lett.}, B486:418--425, 2000.

\bibitem{Kagan:2009gb}
Kagan~A L and Sokoloff~M D.
\newblock {On Indirect \CP Violation and Implications for \Dz - \Dzb and
  $\PB_{(s)}$ - $\Bbar_{(s)}$ mixing}.
\newblock {\em Phys.Rev.}, D80:076008, 2009.

\bibitem{Asner:2010qj}
Asner~D {\it et al.}~[Heavy Flavor Averaging~Group].
\newblock {Averages of b-hadron, c-hadron, and $\tau$-lepton Properties}.
\newblock 2010.
\newblock online update at http://www.slac.stanford.edu/xorg/hfag/index.html.

\bibitem{Grossman:2006jg}
Grossman Y, Kagan~A L, and Nir Y.
\newblock {New physics and \CP violation in singly Cabibbo suppressed D
  decays}.
\newblock {\em Phys.Rev.}, D75:036008, 2007.

\bibitem{Aubert:2007if}
Aubert~B {\it et al.}~[\babar collaboration].
\newblock {Search for \CP violation in the decays $\Dz\to\Km\Kp$ and
  $\Dz\to\pim\pip$}.
\newblock {\em Phys. Rev. Lett.}, 100:061803, 2008.

\bibitem{Staric:2008rx}
Staric~M {\it et al.}~[\belle collaboration].
\newblock {Measurement of \CP asymmetry in Cabibbo suppressed \Dz decays}.
\newblock {\em Phys. Lett.}, B670:190--195, 2008.

\bibitem{Aaltonen:2011zzz}
Aaltonen~T {\it et al.}~[\cdf collaboration].
\newblock {Measurement of \CP--violating asymmetries in $\Dz\to\pip\pim$ and
  $\Dz\to\Kp\Km$ decays at \cdf}.
\newblock {\it Preprint hep-ex 1111.5023}.

\end{thebibliography}

\end{document}